# Spatio-temporal breather dynamics in microcomb soliton crystals

Futai Hu[1,2†*], Abhinav Kumar Vinod[1†*], Wenting Wang[1], Hsiao-Hsuan Chin[1], James F. McMillan[1], Ziyu Zhan[2], Yuan Meng[2], Mali Gong[2], and Chee Wei Wong[1*]

[1] *Fang Lu Mesoscopic Optics and Quantum Electronics Laboratory, University of California, Los Angeles, CA, USA.*

[2] *State Key Laboratory of Precision Measurement Technology and Instruments, Tsinghua University, Beijing 100084, China.*

[†] *These authors contributed equally to this work.*

[*] *Corresponding authors: phyhft@gmail.com, abhinavkumar@ucla.edu, cheewei.wong@ucla.edu*

**Solitons, the distinct balance between nonlinearity and dispersion, provide a route toward ultrafast electromagnetic pulse shaping, high-harmonic generation, real-time image processing, and RF photonic communications. Here we newly explore and observe the spatio-temporal breather dynamics of optical soliton crystals in frequency microcombs, examining spatial breathers, chaos transitions, and dynamical deterministic switching – in nonlinear measurements and theory. To understand the breather solitons, we describe their dynamical routes and two example transitional maps of the ensemble spatial breathers, with and without chaos initiation. We elucidate the physical mechanisms of the breather dynamics in the soliton crystal microcombs, in the interaction plane limit cycles and in the domain-wall understanding with parity symmetry breaking from third-order dispersion. We present maps of the accessible nonlinear regions, the breather frequency dependences on third-order dispersion and avoided-mode crossing strengths, and the transition between the collective breather spatio-temporal states. Our range of measurements matches well with our first-principles theory and nonlinear modeling. To image these soliton ensembles and their breathers, we further constructed panoramic temporal imaging for simultaneous fast- and slow-axis two-dimensional mapping of the breathers. In the phase-differential sampling, we present two-dimensional evolution maps of soliton crystal breathers, including with defects, in both stable breathers and breathers with drift. Our fundamental studies contribute to the**

understanding of nonlinear dynamics in soliton crystal complexes, their spatio-temporal dependences, and their stability-existence zones.

**Introduction**

Self-organization and consequent collective dynamics in coupled dynamical systems are central studies to understand the fundamental mechanism in various fields of our world, such as self-assembled materials and systems[1], biological oscillators[2], transport phenomenon[3], social networks[4], and neural networks[5]. The appearance of nonlinearity and dissipation have a significant impact on the stability, dynamical behavior, and phase transitions in these complex systems[6-8]. As a unique nonlinear object, dissipative solitons[9-14] are wave packets that maintain a double balance of nonlinearity by dispersion and dissipation by gain, providing an excellent route to study nonlinear many-body interactions in a dissipative system. Optical solitons in microresonators have recently attracted tremendous research interest in fields ranging from high-speed optical communications[15], photonic signal processing[16], low-noise radiofrequency generation[17], and coherent distance ranging[18]. The study of soliton stability and dynamics, which is of essential importance in promoting relevant applications, has uncovered various states of a few solitons, including intensity breather solitons[19-22], soliton binding[23-26], and soliton diffusion[27]. In the aspect of collective patterns, the soliton complexes can take the form of crystals in microresonators, with the solitons as constituent elements in self-organized regular patterns analogous to atomic crystals[28-32]. These states can spontaneously form in the presence of a modulated background wave that periodically traps solitons[29,33]. Perfect soliton crystals (PSCs) and soliton crystals with defects have also been observed and characterized[29-31,33]. Many-soliton interactions introduce a new dimension in the control and understanding of soliton states via the interplay between solitons. However, the study of soliton dynamics has hitherto been primarily confined to intensity breathing and slow changes such as melting, recrystallization, and indirect switching of PSCs in the power threshold and frequency detuning subspaces[33]. While spatio-temporal dynamics has been observed in mode-locked fiber resonators[12,31], the breather dynamics of soliton crystals remains largely unexplored in fiber resonators. The energy exchange between soliton tails with background waves in lattice structures has been theoretically exploration[37], hinting at the feasibility of collective motions in soliton crystals. Furthermore, soliton crystals, especially soliton crystals with defects, show advantages in stability, spectral flexibility, and conversion efficiency[34] compared to few solitons, and therefore is being widely examined in recent years[35-38]. However, detailed research on the deterministic generation and spatio-temporal stability of soliton crystals with defects remain elusive.

Here we first examine the breather dynamics of optical solitons in nonlinear 64.8 GHz frequency microcombs, newly including spatially breathers in soliton crystals, phase transitions with perfect crystals and chaotic waveforms. To unravel these nonlinear dynamics, secondly we demonstrate experimentally their spatio-temporal generation pathways via two example transitional maps of the ensemble two-defect and single-defect breathers. With the breathers at tens to hundreds of MHz, we observe transitions involving perfect soliton crystals, chaos, primary comb lines, deterministic crystal $N$-switching from 46× to 48× the free spectral range, and stationary crystal states. Third, we describe the physical mechanisms of the breather spatial dynamics in soliton crystal microcombs, initiated by a asymmetric soliton tail due to emitted dispersive wave and with energy exchange with a background wave. Tracing a limit cycle in the interaction plane, we introduce a collective figure-of-merit for the ensemble motion, including an intracavity noise deviation that increases with the velocity of the collective soliton crystal. The motion of the dark and bright defect-solitons across the soliton crystal is described via two domain-walls with parity symmetry breaking assisted by third-order dispersion. Our nonlinear numerical modeling matches our measurements, presenting maps of the experimentally accessible nonlinear regions and the breather dependences on the third-order dispersion and avoided-mode crossing strengths. Fourth, we imaged these spatio-temporal soliton complexes in real-time via panoramic temporal imaging, watching simultaneously the fast-axis intracavity and slow-axis roundtrip evolutions. With phase-differential sampling, we present two-dimensional evolution maps of the soliton crystals and their breathers, ranging from stable breathers to drifting-breathers to chaotic solitons.

**Results**

**Soliton crystals, breathers, and defects**

Figures 1a-1f show the selected patterns and dynamics of soliton crystals, modeled from the nonlinear Lugiato-Lefever equation for spontaneous pattern formation. Perfect soliton crystals (SCs) are fully occupied solitons in the spatial dimension, as shown in Fig. 1a. In soliton crystals with defects, one may have vacancy, spatially-moving solitons, or interstitial solitons. In Fig. 1b, we illustrate the soliton crystal with one vacancy defect (1-defect SC). Solitons near the defect may diffuse towards the vacancy, subsequently occupying the previous vacancy and leaving behind a new one. Distinct from conventional intensity breathing[21,31,39,40], this process cascades to form a *spatial* breathing behavior as shown in Fig. 1c. Compared to the random diffusion process, the spatial motion of solitons in Fig. 1c is a lasting dynamical process with several to hundreds of MHz repetition rate. The vacancies can be understood as dark defects, while the interstitial solitons can be interpreted as bright defects. These bright defects translate with respect to the soliton crystal and collide elastic-like with stationary solitons. This process can also cascade to form a spatial breather shown in Fig. 1d. Some solitons may chaotically participate in the spatial dynamics while embedded in an 'approximately periodic' structure as depicted in Fig. 1e. The chaotic motion and irregular intensity fluctuations of solitons, sometimes termed as chaotic solitons[41-44], have been reported in different platforms. In this work, we employ the third-order dispersion term to strengthen the parity symmetry breaking near the vacancy, enabling spatial breathing and chaotic motions. Fig. 1f illustrates an example of the switching transition between soliton crystal states of two different spacings $a_1$ and $a_2$, arising from the thermal dependence of the avoided-mode crossing which is detailed in Supplementary Note 1.

**Generation of a family of soliton crystals**

Here we study soliton crystals in a nonlinear Kerr microresonator platform to explore analogous patterns and dynamics. The motion of solitons is determined by the environmental potential provided by both the background wave and waveform tails of other solitons, which can be controlled by the detuning and power. Solitons can be fixed in or escape from the local potential minima. The trapping and escape of solitons have been recently reported in different resonator platforms[45-47]. Double solitons may form soliton molecules through locking to potential minima assisted by dispersive waves[33]. Solitons may also be trapped by periodic potential minima to form a crystallized structure. The interference of two or more strong spectral lines in microresonators can build a periodically modulated background wave that provides such a periodic sequence of potential wells. One of the two spectral lines is usually the coupled-in pump line, and the other lines can be provided by auxiliary lines or enhanced lines via different mechanisms[24,29]. In this work, we use the strong coupling between two eigenmodes and the resulting avoided-mode crossing (AMX) to introduce spectrally local perturbations in several resonances.

We generate various soliton crystals in $Si_3N_4$ Kerr microresonators with measured $Q$ in excess of $10^6$, anomalous dispersion, and free spectral range (FSRs) ≈ 64.8 GHz. With details of the measured cold cavity dispersion noted in Supplementary Note 1, the measured dispersion $D_2/2\pi$ is ≈ 267 kHz. Fig. 2a depicts our experimental setup. A continuous-wave laser followed by an erbium-doped fiber amplifier provides the L-band (long-wavelength band) pump. The inset of Fig. 2a is an optical micrograph of our microresonator. The pump input-output coupling is via bus waveguides and free-space lenses, with two circulators for unidirectional light propagation and two polarizers to purify the light polarization. The pump undergoes an ≈ 1 dB attenuation in the leading circulator and an ≈ 3 dB coupling loss into the bus waveguide. The chip output is separated by a 1×4 fiber coupler, simultaneously measuring the optical and radio-frequency spectra, cavity response, and output power with a piezoelectrically-tuned forward-swept laser (i.e., cavity blue-side to red-side wavelength sweep). The pump line is removed by a 7.5-nm band stop filter before

RF characterization, and a vector network analyzer measures the cavity response with radio-frequency (RF) modulation applied onto the pump via an electro-optic phase modulator.

We experimentally obtain several deterministic paths to access different soliton crystals in a range of devices; these states are then distinguished via numerical simulations and ultrafast temporal imaging. To elucidate the operating map, Fig. 2b shows the modeled spatio-temporal nonlinear dynamics and generation of the soliton crystals via a perturbed Lugiato–Lefever equation[29,33,48,49] (LLE) without temperature independence for illustration clarity. With our experimentally measured parameters, the nonlinear modeling reliably produces the soliton crystals via a forward blue-to-red sweep, with a family of spatio-temporal dynamics over six different regions. As shown in Fig. 2b, PSCs and spatial breathers are generated following chaotic regions, and the spatial breather motion gradually evolves into a stationary state. Fig. 2b further illustrates soliton crystals with one defect (1-defect SCs) and the spatial breather with one defect (1-defect breather). In our nonlinear simulations, stationary SCs and spatial breathers with more than one defect are also observed.

Fig. 2c shows the measured spectral evolution of the microcombs generated at 1593.1 nm with blue-to-red detuning sweep, at 27.0 dBm pump. The soliton crystal generation has five qualitatively different regions, supported by simulation spectra, as: (*i*) primary spectral lines, i.e., temporal Turing patterns, (*ii*) the chaotic waveforms which subsequently lead to the SCs, (*iii*) a 2-defect breather, (*iv*) a 1-defect breather, and (*v*) the stationary multi-defect SCs. We note that the experimental range is slightly larger than the simulated existence frequency ranges due to the resonance shift caused by pump-induced cavity heating in our experiments. The similar intracavity power of chaotic waveforms in region *ii* and breathers in region *iii* makes it thermally stable to reversibly switch between chaotic waveforms and soliton crystals. Aside from the B peak, the VNA responses of breather and chaos states present similarities in Fig. 2e, further reflecting the feasibility of reversible tuning between these two states. This indicates that two states exist at a similar detuning and intracavity power, which is indeed what we observe experimentally, as shown in the trace in Fig. 2c. The formation of breathers in region *iii* is highly repeatable, indicating a deterministic generation pathway. We attribute this to the existence of multiple AMX resonances that contributes to a background potential allowing only specific states. The prominent comb lines in region *v* of Fig. 2c are spaced by $N = 47$ FSRs, denoting a lattice constant of 1/47 roundtrip. A second deterministic state evolution path is shown in Fig. 2d, with a 25.6 dBm pump at 1593.7 nm. This path also covers several sequential regions. We highlight the reversible *N*-transition that alters the lattice constant, wherein the transition points are marked by the dashed white circles. This transition is explained by noting that the AMX perturbation has a temperature dependence. With forward-frequency sweep, the increased intracavity energy induces a higher local temperature, thereby driving the primary AMX perturbation to shift from $\mu_{AMX} = 46$ to $\mu_{AMX} = 47$ and finally $\mu_{AMX} = 48$ as shown in Fig. 2d (pathway II). The number of potential wells is determined by $\mu_{AMX}$, and it hence increases by one after each $\mu_{AMX}$ transition. Since the total number of solitons is conserved during the thermally-stable transition, therefore a vacancy defect is formed ($\mu_{AMX} = 46$ → 47) when the background wave creates an additional potential well. Subsequently, a newly generated soliton can occupy the vacant potential well ($\mu_{AMX} = 47$) with a further detuning change, as shown in Fig. 2c.

We also note that this recurrence of chaotic states, separated by regions of stability, within one forward detuning sweep has rarely been reported in the study of microcombs. This is seen in Fig. 2c, further detailed as pathway I. Distinct from the sudden jump between 2-defect breather and 1-defect breather and stochastic occurrence of chaotic solitons (the latter represented by the dashed white box) in Fig. 2c, we also experimentally observe the continuous and reversible spectral evolution from 1-defect breather to the chaotic solitons. The intracavity spatial motion of solitons in the 1-defect breather gradually slows with a forward detuning sweep and prior to settling down to a stationary state. We therefore observe two repeatable approaches to obtain the soliton crystals with defects and their corresponding breathers: generated from a leading chaotic region shown in Fig. 2c; and generated from the *N*-transition shown in Fig. 1l and Fig. 2d (pathway II). Additionally, the chaotic solitons itself will be further detailed in the temporal observation section later in this manuscript.

We next characterize the cavity response and breathers shown in Fig. 2c. With pump phase modulation, the cavity response is obtained on the pump line using a vector network analyzer[31]. Fig. 2e plots the vector network analyzer magnitude for the four different states of Fig. 2c. We observe two major peaks in the vector network analyzer spectra for the 2-defect SC, namely the cavity (*C*)-resonance and soliton (*S*)-resonance. *C*-resonance reflects the effective cavity resonance considering the frequency shift due to cross-phase modulation from the background wave on the phase modulation sidebands. *S*-resonance reflects the effective soliton resonance that deviates from the *C*-resonance due to cross-phase modulation mainly from solitons[23]. The *S*-resonance, together with the optical spectrum, confirms the existence of soliton crystals. The strength of *S*-resonance is dependent on the number of solitons within the cavity. Since we have 45 surviving solitons in a 2-defect SC, the corresponding *S*-resonance is much stronger than *C*-resonance. As the SCs approach the breather states, the *S*- and *C*-resonances become closer and difficult to distinguish. Furthermore, there exists a new peak featured in both the 2-defect and 1-defect breathers induced by the dynamical breathing, which we term the "breathing (*B*)-peak". The *B*-peak in the 2-defect SC is located at almost the same frequency as the 1-defect SC due to the similar breathing frequency. We expect the magnitude of the *B*-peak to be correlated with the number of spatial breather solitons at each instance (which in this study directly corresponds to the number of defects). This is indeed observed in the experiment with the 2-defect SC having a higher *B*-peak than the 1-defect SC. The *B*-peaks shift to a lower frequency with a forward pump sweeping, verifying the slowing of spatial breathers. The *B*-peak is almost missing when SC evolves into a chaotic state.

**Discussion**

**Mechanism of spatial motion in soliton crystals**

Besides the resemblance in form between atomic and soliton crystals, we find that soliton kinetics in soliton crystals can also be explained via the concepts of velocity, potential, and energy, analogous to particle mechanics. Fig. 3a shows the numerical model of a 1-defect breather with a breathing frequency similar to an experimental $f_B \approx$ 60 MHz illustrated in the right inset. The soliton propagation velocity is slightly altered when near the defect compared to the SC background; this soliton thus escapes from the local potential minima and drifts with respect to the crystal ensemble. This drifting soliton occupies the vacant potential well and generates a new defect at its original position. This unpinning and subsequent occupation of defects by adjacent solitons repeats periodically, resulting in a *spatially breathing crystal* and a periodic variation of the intracavity power. Spatial breathing results from the many-body interactions, and the spatial motion is relative to the soliton crystal background. The spatial breathing of solitons thus occurs due to a physically distinct mechanism compared to traditionally observed soliton breathers or the vibration of soliton molecules. When defects and solitons move across each other, remarkably, the solitons remain stable. The spatial breathing might however become unstable during the transition between the breather and stationary state, giving rise to chaotic solitons that are identifiable in their RF spectra, detailed later in Figure 4.

In the left inset of Fig. 3a, we plot the resultant selected intracavity spatial snapshots from the dashed white rectangular region. The first intracavity peaks are aligned to illustrate the soliton dynamics. The dispersive waves induced by third-order dispersion (TOD) lead to asymmetric soliton tails, with higher intensity at the leading edge of the soliton. The emitted dispersive wave by one soliton exchanges energy with the background wave and other solitons. At the beginning of the spatial motion, the peak intensity and phase of this candidate soliton gradually vary. This process imparts the candidate soliton with a lower velocity than the remaining soliton ensemble. The energy consistently transfers from the leading front to the trailing edge as the candidate soliton moves. The moving soliton approaches the vacant potential minima and its group velocity with respect to the crystal decreases gradually. Subsequently, the candidate soliton is trapped in the position of the original defect.

Corresponding to the spatial breather of Fig. 3a right inset, Fig. 3b illustrates the measured optical spectrum of the 1-defect SC, with prominent lines marked in yellow. The AMX and pump resonances are denoted by the purple and red arrows respectively. The spatial waveform of the stationary 1-defect SC can be understood as the destructive interference between PSC and an out-of-phase soliton located in one potential well[31]. Hence, if we filter out the prominent lines (effectively arising from the PSC) of 1-defect SC, the remaining optical waveform would be a soliton-like pulse located at the vacancy. We use the same strategy to understand the 1-defect breather. Fig. 3c depicts the evolution map of the waveform after this filtering. Three pulses labeled with $p_{origin}$, $p_{mid}$, and $p_{new}$ are involved in the spatial motion of one soliton. As shown in Fig. 3c inset, we note that $p_{new}$ is in phase with $p_{origin}$ and out of phase with $p_{mid}$. At the beginning of soliton motion, $p_{new}$ and $p_{mid}$ are close enough to form a deconstructive interference. With the soliton moving, $p_{mid}$ then moves away from $p_{new}$ and towards $p_{origin}$, causing a deconstructive interference with $p_{origin}$. These dynamics indicate the annihilation of the original dark defect and the creation of a new dark defect in 1-defect SC.

The antithesis of the dark defect is the bright defect shown in Fig. 1j, where the moving soliton collides with the next one elastically. The dynamics of the dark and bright defects show unidirectional propagation with respect to the crystal background. The 1-defect crystal can be understood as two domain-walls connecting two PSCs and the defect[29,50-52]. The parity symmetry breaking induced by TOD causes *asymmetric* crystal structure at two sides of the defect, acting as an external force. The externally driven domain-walls propagate unidirectionally, transiting the candidate soliton. The domain-wall is sometimes termed "topological defects" or "topological soliton" which exhibits the topological robustness while moving[53]. Here we compare our simulation results to our experimental data – to compare equivalently, we also include the frequency microcomb noise. As shown in Fig. 3d, we first estimate the intracavity noise via the microcombs power recorded with a high-speed photodetector and real-time oscilloscope with 100 GSa/s sampling rate (photodetector bandwidth at 20 GHz). A 4-nm-bandwidth bandpass filter, illustrated by the dashed purple rectangle in Fig. 3b, is applied to increase the signal contrast. Details of intracavity noise estimation can be found in Supplementary Note 2, including the limit cycle fluctuations. Here all noise effects are attributed to detuning disturbances up to 5 MHz for simplicity. The simulated AC-coupled intensity has a clean curve without considering intracavity noise, while notable noisy experiment-like signatures appear after considering intracavity noise. This intracavity noise varies with the pump power and frequency.

To understand the soliton complexes and associated transport, we parametrized the dynamical trajectories of solitons in the interaction plane[54,55] with complex number $\rho e^{i\varphi}$ where $\rho$ and $\varphi$ represent the spacing and phase difference between subsequent soliton peaks respectively. We further define a collective figure-of-merit for the soliton crystal to describe the collective motion:

$$\boldsymbol{\rho}_{\text{coll}} = \sum_{n=1}^{N-D} \rho_n e^{i\varphi_n} / (N-D) \quad (1)$$

where $N$ is the number of solitons if the crystal were perfect, $D$ is the number of defects, $\rho_n$ and $\varphi_n$ are the spacing and phase difference between $n$th and $(n+1)$th soliton peaks respectively. In Eq.(1), the periodic boundary condition $\rho_n = \rho_{n+N-D}$, $\varphi_n = \varphi_{n+N-D}$ is applied. Fig. 3e illustrates the resulting modulus and phase of $\boldsymbol{\rho}_{\text{coll}}$ from 16,000 roundtrips, depicting intracavity traces along a limit cycle. These traces may slightly deviate from the limit cycle with estimated intracavity noise. Interestingly, the deviation increases with the collective velocity ($v_{\text{coll}} = \partial \boldsymbol{\rho}_{\text{coll}}/\partial t$) as marked by red arrows, where the slow time $t$ increases with roundtrips. In the inset, we also plot the trace of all $\rho_n$ in the 16,000 roundtrips. Most of the time, $\rho_n$ is located near the crossing between two dashed grey lines evidencing a stable and in-phase SC background -- indicating that the solitons move sequentially.

As detailed above, the asymmetric tail of solitons induced by TOD and the background wave governed by AMX largely determine the spatial breathing. As shown in Fig. 3f, the $f_B$-$D_3$ dependence follows the same trend at different detunings and reaches a stationary point beyond $D_3 = -2$ kHz. These curves indicate that the asymmetric tail assisted by the TOD is the decisive cause for the spatial motion of solitons. AMX introduces local perturbations in the modal frequencies of the microresonator, resulting in the formation of primary lines at these locations. Furthermore, once soliton pulses are generated, this modal mismatch due to AMX may result in the formation of dispersive waves[56,57]. Dispersive waves of soliton pulses interlock to form a periodic background potential field. These background field act as potential wells and trap solitons by helping balance the nonlinearity and dispersion locally[35]. The intensity of primary lines as well as dispersive waves depends on the AMX strength. The presence of third order dispersion however creates an asymmetry in the soliton profile which allows solitons to escape in a preferential direction when perturbed. At certain values of detuning and pump power, solitons may periodically enter and exit the empty potential well which leads to spatial breathing. If the dispersive wave caused by AMX has higher intensity, the intracavity maxima are correspondingly stronger and apply a larger binding force on solitons. This explains the dependence between the breathing frequency and AMX strength. This also indicates that solitons may no longer be stably locked in potential wells if the dispersive wave intensity is low.

**Experimentally recorded and simulated Kerr soliton crystals**

Furthering from the dynamics and deterministic generation of different soliton crystals in the above section, here we present a further detailed characterization of optical spectra, RF spectra, and tunable breathing frequencies of the crystal states. The optical spectra of some representative soliton crystals and corresponding RF spectra are summarized in Fig. 4a and Fig. 4b. Most experimental spectra except the 1-defect SC are measured with a 25.5 dBm pump at 1593.7 nm, while the 1-defect SC is measured with a 25.5 dBm pump at 1593.1 nm. Two stationary states, including the PSC and 1-defect SC, have low-noise RF spectra. The purple top arrow marks the dominant AMX point that is spaced by ≈ 48 FSRs with the pump resonance, with the residual background near the pump from amplified spontaneous emission. The numerically modeled reproduction of these two states is shown in Fig. 4c. The breathing frequencies of the 1-defect and 2-defect breathers are about 150 MHz and 137 MHz. The temporal variation of the microcombs power is not a perfect sinusoid, which corresponds to harmonic peaks. We highlight the spectral interference pattern in the 2-defect breather, directly correlated with the spacing between two defects. We determine the defect spacing to be 12/48 roundtrip by comparing this spectral pattern to simulated results in Fig. 4c. The simulated dynamical and stationary structures confirm our hypothesis. The spectral similarity between spatial breathers and stationary SCs is a useful tool to deduce defect spacing. When SCs become chaotic, we observe significant spectral fluctuations compared to the spatial breather. In RF spectra of chaotic solitons, the chaotic pattern appears but surprisingly exhibits some unique features distinct from normal chaotic states. The RF noise is not solely extended to fill low RF frequencies but also distributes around the breathing frequency and its harmonics. We also see that the chaotic motion mixes with the intensity fluctuation of soliton crystals through the modeled nonlinear temporal structure.

Fig. 4e and Fig. 4f plot the RF information of SCs measured in two selected generation paths in Fig. 2c and Fig. 2d. The datasets are recorded with a fast sweep to avoid the slow drift of the pump detuning. The breathing frequency can be tuned within a strikingly broad range from 56 MHz to 128 MHz in Fig. 4e. At a lower pump power of 25.5 dBm at 1593.1 nm, we even get a breathing frequency down to ≈ 10 MHz. Such a broad tuning range has yet been observed in microresonator breather soliton states and is attributed to the unique breathing mechanism. The breathing frequencies generally decrease with a blue-to-red sweep in both generation paths. Interestingly, the slope sign of 1-defect breather in Fig. 4e tends to zero and even flips to be a positive value at low breathing frequencies. The simulation-accessible $f_B$ under 10 MHz is difficult to experimentally access due to the appearance of chaotic waveforms at the low breathing frequency. We attribute this phenomenon to the intracavity RF noise characterized in the above paragraph. Therefore, we summarize two mechanisms to generate chaotic soliton waveforms: the unbalance between the AMX strength and TOD, as discussed in the former section; the detuning disturbance induced by intracavity noise (detailed in Supplementary Note 2).

In Fig. 4g we simulated the detuning dependence of the breathing frequencies at different input powers. The flip of the slope sign is observed at 100 mW, 150 mW, and 200 mW. With increasing input power, the existence range and breathing frequencies of breathers generally red-shift. The 1-defect breather and 2-defect breather share similar existence range and breather frequencies, while they exist in different frequency ranges in Fig. 4e. The number switching of defects in experiments can be explained by the background wave variation due to the thermal dependence of AMX strengths (further detailed in Supplementary Note 1&2).

**Temporal observation of the dynamical breathing**

In prior sections, we have explored the deterministic generation, dynamical properties, and the physical mechanism of different soliton crystals. Soliton crystals have hitherto primarily been studied on slow-time scale[29,54,56] using spectral measurements via an OSA or temporal characterization as measured by intensity auto- or cross-correlation. However, these methods are unable to accurately capture dynamically evolving crystal states, such as breathing SCs or quasi-chaotic solitons. The prior methods used to record the dynamical temporal structure include large bandwidth oscilloscope[31], time-lens[23,57], and time-stretched dispersive Fourier transform (TS-DFT)[58]. In this section, we further delve into the temporal behavior of soliton crystals observed in a panoramic-reconstruction temporal imaging (PARTI) system[54].

Fig. 5a manifests the experimental setup for understanding the spatial breathers. A soliton sequence with an angular breathing frequency $\omega_B$ is generated from DUT and recorded separately in slow- and fast-time scales. Our temporal imaging system primarily consists of a four-wave mixing (FWM) based time lens and an optical buffer to increase the record length of a single frame[58]. The generated breathing soliton crystal first goes through BPF2 to match the FWM bandwidth. The maximum record length for one shot is limited to around 500 ps. We effectively extend the recording length in one frame to 2 ns by generating 9 replicas in the optical buffer and stitching their images with a fixed temporal shift. The frame rate of our temporal imaging system is 2 MHz.

The repetition rate of the breathers reaches over 60 MHz and the period of the breathers is much longer than the PARTI frame length of 2 ns. The detailed breather dynamics is too fast to be captured using direct photoelectrical detection and the period is too long to be captured in one frame in PARTI. The phase-differential sampling we use is inspired by the stroboscopic effect. Fig. 5b shows the phase-differential sampling schematic of using a low frame frequency of $f_F$ to capture the full dynamics of a periodic signal with a high repetition rate $f_B$. We can use the stroboscopic effect to stitch the dynamics of breathers in a full period. $\phi_{res}$ is the accumulated phase difference between the SC breather and time-lens sampling after 500 ns. Here the blue circle represents the frame repetition in the temporal imaging system while the red circle represents the periodic signal repetition. The residual phase difference between the periodic signal and the frame after $\tau_F = 1/f_F$ is given by $\varphi_{res} = 2\pi \ mod \ (f_B / f_F)$, where ***mod*** is the remainder function. The frame length $\tau_L \approx 2$ ns can also be expressed in the form of a phase span of $\varphi_L = 2\pi f_B \tau_L$. Considering $\varphi_{res}$ in the range of (0, π), the dynamics of the signal within one period is down-sampled when $\varphi_{res} > \varphi_L$ and over-sampled when $\varphi_{res} < \varphi_L$. The generated soliton crystal first goes through BPF2 to match the bandwidth of FWM. The simulated waveform of 2-defect SC in one roundtrip after BPF2 is plotted in Fig. 5c. Although the crystal-like structure of the soliton ensemble is masked due to the limited FWM and BPF2 spectral bandwidth, the spectral bandwidth is still sufficient to resolve the spatial dips corresponding to the soliton crystal defects. Fig. 5d exhibits the numerically modeled of the breather crystal dynamics over large roundtrips. The spectrally-filtered trace of the 2-defect breathers shows that spatial dips occur with a constant period marked by the white dashed line. The spatial spacing of two dips is presented via a black dashed line. The sequent generation of two dips indicates that the solitons near two defects spatially move with a temporal phase difference, as outlined by the purple dashed line. Using a 2 MHz sampling rate and a 2 ns frame length, Fig. 5e shows the simulated 2D trace through the PARTI metrology system. The resampled map reconstructs two dips via resampling and stitching multiple breathing periods. This process is in the same principle as the stroboscopic effect.

Fig. 5f shows the experimental 2D evolution map of the 2-defect breather recorded by the PARTI temporal imaging. We select a crystal state with a breathing frequency of ≈ 56 MHz, corresponding to an ≈ 18 ns temporal period. Therefore, we use nine frames or more to fully recover the breathing dynamics. In parallel to the simulated evolution map, we clearly illustrate the resampled period, spatial spacing and breathing phase difference of two dips in Fig. 5f, reflecting that the solitons near the two defects move with a non-zero phase offset. The breathing frequency is quite stable, inducing a nearly periodic pattern.

In contrast, the breathing frequency of the 2-defect breather drifts sizably in Fig. 5g, giving rise to a decreasing period of the dips with the number of roundtrips which indicates that $\varphi_{res}$ varies away from 0. Fig. 5h, with its finite-linewidth breathing peak, verifies the breathing nature of the SC, including drift. When chaotic waveforms occur, the clear evolution shown in Fig. 5f and Fig. 5g is replaced by a chaotic process with dips randomly appearing as shown in Fig. 5i. The inset shows the corresponding RF spectrum, with a clear low-frequency noise for the chaotic state of the multi-soliton nonlinear dynamics.

**Conclusion**

In summary, we observe the formation and evolution of various soliton crystal states, including the presence of different kinds of defects and breather states. We show that the soliton crystal dynamics is caused by AMX in silicon nitride microresonators with suitable modal dispersion. We explored the generation pathways of these crystal states and observed the first occurrence of spatial soliton crystal breathers in microresonators which are distinct from traditionally observed soliton intensity breathers due to their periodic spatial motion with respect to the remaining solitons within the crystal. We subsequently established different pathways to reliably generate these breather soliton crystals and explored transitions between the number of defects and soliton number, both induced by the thermal dependence of the AMX. We next described the formation of these states in simulations and elucidated the mechanisms of spatial breathing and the existence range of these crystal breathers. We further correlated both in experiments and simulations that the soliton breathing frequency is tunable continuously from several MHz to a few hundred MHz by changing pump detuning. We subsequently mapped the spatial breathing in real-time by observing the evolution of the defect within the crystal via our nonlinear time-lens system. This study explores the dynamics of soliton crystal breathers that have hitherto been relatively unexplored and contribute to advances in nonlinear dynamics, many-body physics and practical applications of soliton crystals[58]. Our study helps the modal dispersion design of microresonators used for soliton crystals, to avoid possible unwanted states. We present several deterministic paths to generate soliton crystals with defects, especially the path to soliton crystals with one defect and perfect soliton crystal. Soliton crystals with one defect have been used for dense data communication[40], radiofrequency signal processing[63,] and optical neuromorphic processing[5].

**Methods**

**Device nanofabrication.** We designed a high-$Q$ $Si_3N_4$ microresonator, nanofabricated via a CMOS-compatible process[59,60]. The fabrication process is as follows: we first deposit a 5 µm-thick $SiO_2$ bottom oxide via plasma-enhanced CVD on a silicon wafer which acts as the under-cladding to minimize loss to the substrate. We then deposit 800 nm $Si_3N_4$ via LPCVD for the resonator structures. This layer is then patterned with deep-ultraviolet lithography and etched down to the cladding oxide via reactive ion dry etch. The wafer is then annealed at 1150°C, to reduce the prevalence of N-H, Si-H, and Si-O-H bonds and thereby reduce propagation loss. The structures are then capped with a 3 µm top oxide.

The resonator consists of two tapered straight waveguides connected by semicircular regions, one of which is coupled to the input bus. This design is chosen to maintain cavity $Q$ while simultaneously minimizing the presence of multiple mode-crossing points and achieving anomalous dispersion around the pump wavelength. The 200 µm bend radius and the 1 µm waveguide width suppress higher-order modes. The straight waveguides are tapered from 1 µm at the edge to 2 µm at the center and back. This waveguide geometry allows us to achieve the targeted cavity dispersion and filter higher-order modes. The adiabatic nature of the taper ensures the preservation of the cavity $Q$.

**Processing of PARTI datasets.** The datasets recorded by the PARTI system include frames with a 2 MHz repetition rate. Fig. 6a plots one of the frames corresponding to the measurement maps shown in Fig. 5f and Fig. 5g. Each frame contains one blank region and ten replicas that cascade with a constant time delay. The blank region is designed to distinguish different frames. Fig. 6b plots the zoom-in of one replica marked by red dashed rectangular in Fig. 6a. There are several roundtrips within one replica, and we depict one example roundtrip in the Fig. 6b inset. During the data processing, we first split each replica into roundtrips, and then we use ($V$-$V_{min}$)/ ($V$-$V_{max}$) to normalize the recorded intensity. The roundtrip time is manually selected to clearly illustrate the evolution of the recorded intensity. Here we set the minimum voltage in each roundtrip to be zero, thereby increasing the contrast between two out-of-phase dips shown in Fig. 5f. We delete several roundtrips at the beginning and end of each replica to remove the redundant roundtrips that overlap on subsequent and prior replicas and thereby increase signal quality. Subsequently, we stitch roundtrips together in each replica and stitch replicas together in each frame. The blank region in each frame is deleted. Then we stitch all frames together to obtain the evolution maps in Fig. 5f and Fig. 5g.

**References**

1  Versluis, F., van Esch, J. H. & Eelkema, R. Synthetic self-assembled materials in biological environments. *Adv Mater* **28**, 4576-4592, (2016).
2  Rubenstein, M., Cornejo, A. & Nagpal, R. Programmable self-assembly in a thousand-robot swarm. *Science* **345**, 795-799, (2014).
3  Strogatz, S. H. & Stewart, I. Coupled Oscillators and Biological Synchronization. *Sci Am* **269**, 102-109, (1993).
4  Hernandez-Ortiz, J. P., Stoltz, C. G. & Graham, M. D. Transport and collective dynamics in suspensions of confined swimming particles. *Phys Rev Lett* **95**, 204501, (2005).
5  Watts, D. J. & Strogatz, S. H. Collective dynamics of 'small-world' networks. *Nature* **393**, 440-442, (1998).
6  Timme, M. & Wolf, F. The simplest problem in the collective dynamics of neural networks: is synchrony stable? *Nonlinearity* **21**, 1579-1599, (2008).
7  Casadiego, J., Nitzan, M., Hallerberg, S. & Timme, M. Model-free inference of direct network interactions from nonlinear collective dynamics. *Nat Commun* **8**, 2192 (2017).

8   Desai, R. C. & Kapral, R. *Dynamics of Self-organized and Self-assembled Structures*. (Cambridge University Press, 2009).
9   Witthaut, D. *et al.* Collective nonlinear dynamics and self-organization in decentralized power grids. *Rev Mod Phys* **94**, 015005, (2022).
10  Stegeman, G. I. & Segev, M. Optical spatial solitons and their interactions: Universality and diversity. *Science* **286**, 1518-1523, (1999).
11  Buryak, A. V., Di Trapani, P., Skryabin, D. V. & Trillo, S. Optical solitons due to quadratic nonlinearities: from basic physics to futuristic applications. *Phys Rep* **370**, 63-235, (2002).
12  Grelu, P. & Akhmediev, N. Dissipative solitons for mode-locked lasers. *Nat Photonics* **6**, 84-92, (2012).
13  Xiao, Z. *et al.* Near-zero-dispersion soliton and broadband modulational instability Kerr microcombs in anomalous dispersion. *Light: Science & Applications* **12**, 33, (2023).
14  Li, Z. *et al.* Ultrashort dissipative Raman solitons in Kerr resonators driven with phase-coherent optical pulses. *Nat Photonics*, 1-8, (2023).
15  Herr, T. *et al.* Temporal solitons in optical microresonators. *Nat Photonics* **8**, 145-152, (2014).
16  Marin-Palomo, P. *et al.* Microresonator-based solitons for massively parallel coherent optical communications. *Nature* **546**, 274-279, (2017).
17  Hu, J. Q. *et al.* Reconfigurable radiofrequency filters based on versatile soliton microcombs. *Nat Commun* **11**, 4377, (2020).
18  Lucas, E. *et al.* Ultralow-noise photonic microwave synthesis using a soliton microcomb-based transfer oscillator. *Nat Commun* **11**, 374, (2020).
19  Riemensberger, J. *et al.* Massively parallel coherent laser ranging using a soliton microcomb. *Nature* **581**, 164-170, (2020).
20  Dudley, J. M., Dias, F., Erkintalo, M. & Genty, G. Instabilities, breathers and rogue waves in optics. *Nat Photonics* **8**, 755-764, (2014).
21  Yu, M. J. *et al.* Breather soliton dynamics in microresonators. *Nat Commun* **8**, 14569, (2017).
22  Xu, G., Gelash, A., Chabchoub, A., Zakharov, V. & Kibler, B. Breather Wave Molecules. *Phys Rev Lett* **122**, 084101, (2019).
23  Lucas, E., Karpov, M., Guo, H., Gorodetsky, M. L. & Kippenberg, T. J. Breathing dissipative solitons in optical microresonators. *Nat Commun* **8**, 736, (2017).
24  Weng, W. L. *et al.* Heteronuclear soliton molecules in optical microresonators. *Nat Commun* **11**, 2402, (2020).
25  Guo, H. *et al.* Universal dynamics and deterministic switching of dissipative Kerr solitons in optical microresonators. *Nat. Phys.* **13**, 94-102, (2017).
26  Liu, Y. *et al.* Phase-tailored assembly and encoding of dissipative soliton molecules. *Light: Science & Applications* **12**, 123, (2023).

27  Turaev, D., Vladimirov, A. G. & Zelik, S. Long-Range Interaction and Synchronization of Oscillating Dissipative Solitons. *Phys Rev Lett* **108**, 263906, (2012).
28  Bao, C. Y. *et al.* Quantum diffusion of microcavity solitons. *Nat. Phys.* **17**, 462–466, (2021).
29  Cole, D. C., Lamb, E. S., Del'Haye, P., Diddams, S. A. & Papp, S. B. Soliton crystals in Kerr resonators. *Nat Photonics* **11**, 671-676, (2017).
30  Yao, B. C. *et al.* Gate-tunable frequency combs in graphene-nitride microresonators. *Nature* **558**, 410–414, (2018).
31  Karpov, M. *et al.* Dynamics of soliton crystals in optical microresonators. *Nat. Phys.* **15**, 1071-1077, (2019).
32  Nie, M. *et al.* Dissipative soliton generation and real-time dynamics in microresonator-filtered fiber lasers. *Light: Science & Applications* **11**, 296, (2022).
33  Lu, Z. Z. *et al.* Synthesized soliton crystals. *Nat Commun* **12**, 3179, (2021).
34  Andrianov, A. V. All-Optical Manipulation of Elastic Soliton Crystals in a Mode-Locked Fiber Laser. *Ieee Photonic Tech L* **34**, 39-42, (2022).
35  Boggio, J. M. C. *et al.* Efficient Kerr soliton comb generation in micro-resonator with interferometric back-coupling. *Nat Commun* **13**, 1292, (2022).
36  Corcoran, B. *et al.* Ultra-dense optical data transmission over standard fibre with a single chip source. *Nat Commun* **11**, (2020).
37  Tan, M., Xu, X. & Moss, D. J. Optical neuromorphic processing at Tera-OP/s speeds based on Kerr soliton crystal microcombs. *arXiv preprint arXiv:2105.06296*, (2021).
38  Moss, D. Real-Time Video Image Processing based on Kerr Soliton Krystal Microcombs. (2021).
39  Taheri, H., Matsko, A. B., Maleki, L. & Sacha, K. All-optical dissipative discrete time crystals. *Nat Commun* **13**, 848, (2022).
40  Mussot, A. *et al.* Fibre multi-wave mixing combs reveal the broken symmetry of Fermi-Pasta-Ulam recurrence. *Nat Photonics* **12**, 303, (2018).
41  Vinod, A. K. *et al.* Frequency microcomb stabilization via dual-microwave control. *Commun Phys-Uk* **4**, 81, (2021).
42  Akhmediev, N., Soto-Crespo, J. M. & Town, G. Pulsating solitons, chaotic solitons, period doubling, and pulse coexistence in mode-locked lasers: Complex Ginzburg-Landau equation approach. *Phys Rev E* **63**, 056602, (2001).
43  Ustinov, A. B., Demidov, V. E., Kondrashov, A. V., Kalinikos, B. A. & Demokritov, S. O. Observation of the chaotic spin-wave soliton trains in magnetic films. *Phys Rev Lett* **106**, 017201, (2011).
44  Wei, Z. W. *et al.* Pulsating soliton with chaotic behavior in a fiber laser. *Opt Lett* **43**, 5965-5968, (2018).
45  Xin, F. F. *et al.* Evidence of chaotic dynamics in three-soliton collisions. *Phys Rev Lett* **127**, 133901, (2021).

46   Jang, J. K., Erkintalo, M., Coen, S. & Murdoch, S. G. Temporal tweezing of light through the trapping and manipulation of temporal cavity solitons. *Nat Commun* **6**, 7370, (2015).
47   Garbin, B., Javaloyes, J., Tissoni, G. & Barland, S. Hopping and emergent dynamics of optical localized states in a trapping potential. *Chaos* **30**, 093126, (2020).
48   Chembo, Y. K. & Menyuk, C. R. Spatiotemporal Lugiato-Lefever formalism for Kerr-comb generation in whispering-gallery-mode resonators. *Phys Rev A* **87**, 053852, (2013).
49   Godey, C., Balakireva, I. V., Coillet, A. & Chembo, Y. K. Stability analysis of the spatiotemporal Lugiato-Lefever model for Kerr optical frequency combs in the anomalous and normal dispersion regimes. *Phys Rev A* **89**, 063814, (2014).
50   Parra-Rivas, P., Gomila, D., Gelens, L. & Knobloch, E. Bifurcation structure of localized states in the Lugiato-Lefever equation with anomalous dispersion. *Phys Rev E* **97**, 042204, (2018).
51   Wang, H. *et al.* Self-regulating soliton domain walls in microresonators. *arXiv preprint arXiv:2103.10422*, (2021).
52   Parra-Rivas, P., Coulibaly, S., Clerc, M. G. & Tlidi, M. Influence of stimulated Raman scattering on Kerr domain walls and localized structures. *Phys Rev A* **103**, 013507, (2021).
53   Parra-Rivas, P., Gelens, L., Hansson, T., Wabnitz, S. & Leo, F. Frequency comb generation through the locking of domain walls in doubly resonant dispersive optical parametric oscillators. *Opt Lett* **44**, 2004-2007, (2019).
54   Herink, G., Kurtz, F., Jalali, B., Solli, D. R. & Ropers, C. Real-time spectral interferometry probes the internal dynamics of femtosecond soliton molecules. *Science* **356**, 50-53, (2017).
55   Soto-Crespo, J. M., Grelu, P., Akhmediev, N. & Devine, N. Soliton complexes in dissipative systems: Vibrating, shaking, and mixed soliton pairs. *Phys Rev E* **75**, 016613, (2007).
56   Wang, W. Q. *et al.* Robust soliton crystals in a thermally controlled microresonator. *Opt Lett* **43**, 2002-2005, (2018).
57   Li, Y. N. *et al.* Real-time transition dynamics and stability of chip-scale dispersion-managed frequency microcombs. *Light-Sci Appl* **9**, 52, (2020).
58   Li, B. W., Huang, S. W., Li, Y. N., Wong, C. W. & Wong, K. K. Y. Panoramic-reconstruction temporal imaging for seamless measurements of slowly-evolved femtosecond pulse dynamics. *Nat Commun* **8**, 61, (2017).
59   Xu, X. Y. *et al.* Broadband photonic RF channelizer with 92 channels based on a soliton crystal microcomb. *J Lightwave Technol* **38**, 5116-5121, (2020).
60   Huang, S. W. *et al.* Smooth and flat phase-locked Kerr frequency comb generation by higher order mode suppression. *Sci Rep-Uk* **6**, 26255, (2016).

**Acknowledgments**

The authors acknowledge discussions with Peiqi Wang, Yuanmu Yang, Hao Liu, Tristan Melton, Jiagui Wu, Hsiao-Hsuan Chin, Dong-Il Lee, Allen Kuan-Chen Chu, Alwaleed Aldhafeeri, and Alex Wenxu Gu. The authors acknowledge the device nanofabrication from the Institute of Microelectronics in Singapore, and the microresonator layouts from Jinghui Yang. The support provided by the China Scholarship Council (CSC) during a visit of Futai Hu to UCLA is acknowledged. This work is supported by the National Science Foundation, DARPA, and Lawrence-Livermore National Laboratory.

**Author contributions**

F.H., A.K.V., W.W., and J.F.M. performed the measurements; F.H., A.K.V., Z.Z., and M.G. performed the measured data analysis and contributed to the physical explanation on collective dynamics. F.H., A.K.V. and Y.M. performed the nonlinear simulations and modeled data analysis. F.H., A.K.V., and C.W.W wrote the manuscript with contributions from all authors. C.W.W. supervised and supported this research.

**Competing interests**

The authors declare no competing interests.

**Supplementary Information**

The online version contains supplementary material available. All data needed to evaluate the conclusions in the paper are present in the paper and available. Raw data used to produce the figures in the main text and supplementary materials are available upon request.

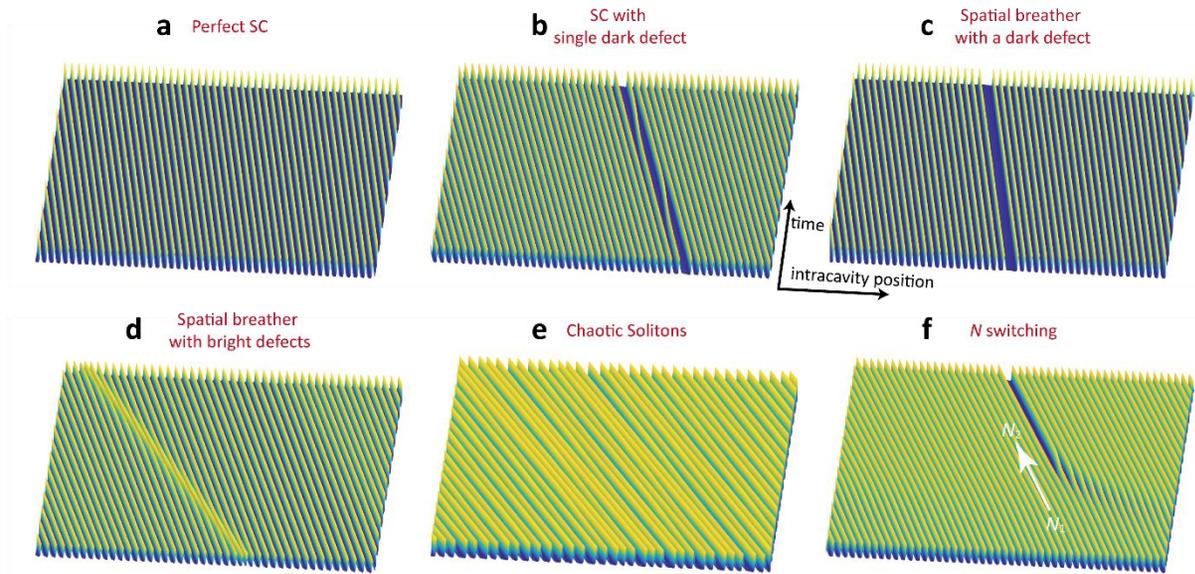

**Figure 1 | Breather dynamics of microcomb soliton crystals | Schematic illustration of the soliton crystals, breathers and dynamics.** (**a** to **f**) Spatial-temporal structure of different optical soliton crystals (SCs), in correspondence with atomic crystals. **a,** Perfect SC, in periodic potential wells from the extended background electromagnetic wave. **v,** SC with single dark defect. The soliton in periodic crystals can be individually missing, leaving a dark vacancy defect. **c,** Spatial breather with one dark defect. The soliton near the defect can diffuse to the other side of the defect. **d**, Spatial breather with one bright defect. The interstitial soliton interacts with the neighboring soliton with an elastic-like collision. **e**, Chaotic SC, with fluctuating motions of soliton crystals including intensity variations. **f**, $N$-switching of soliton crystal, with a new potential well generated without a soliton occupying, as well as the addition of a vacancy.

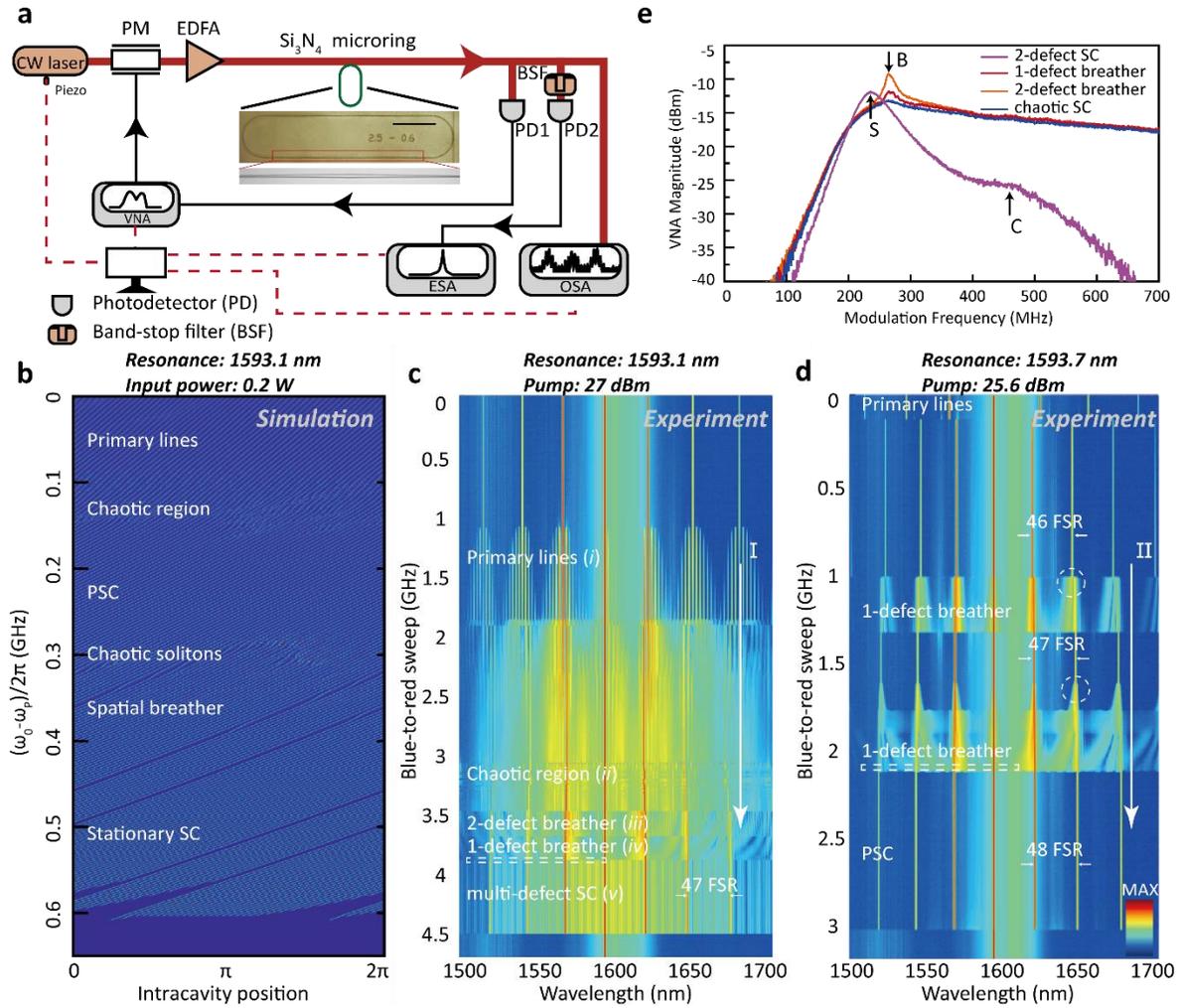

**Figure 2 | Experimental generation of soliton crystal (SC) microcombs with different dynamical regions. a**, Experimental setup for soliton crystal generation. A continuous-wave (CW) laser drives the crystal generation with the wavelength tuned piezoelectrically. PM: phase modulator, EDFA: erbium-doped fiber amplifier, VNA: vector network analyzer, ESA: electronic spectrum analyzer, and OSA: optical spectrum analyzer. Here the ESA resolution bandwidth is 100 kHz. Inset: Microscope image of device under test. Scale bar: 200 μm. **b**, The generation path of soliton crystals, from nonlinear numerical modeling. We identify six distinct states including primary lines, chaotic states, PSC, chaotic SC, spatial breathers, and stationary SC. (**c** and **d**), Experimental optical spectra of generated microcombs from a 64.8 GHz $Si_3N_4$ microresonator using forward (blue-to-red) laser sweeping at 27.0 dBm (**c**) and 25.6 dBm (**d**) input powers respectively. The white dashed rectangle denotes the region where chaotic soliton crystals stochastically occur. The white dashed circle marks the region where the spacing between adjacent strong spectral lines changes from 46 FSRs to 47 FSRs and subsequently 48 FSRs. **e**, The VNA response plotted against modulation frequency in different SC states. "*B*" represents the peak induced by spatial breathing. "*S*" and "*C*" represents the *S*- and *C*-resonances respectively.

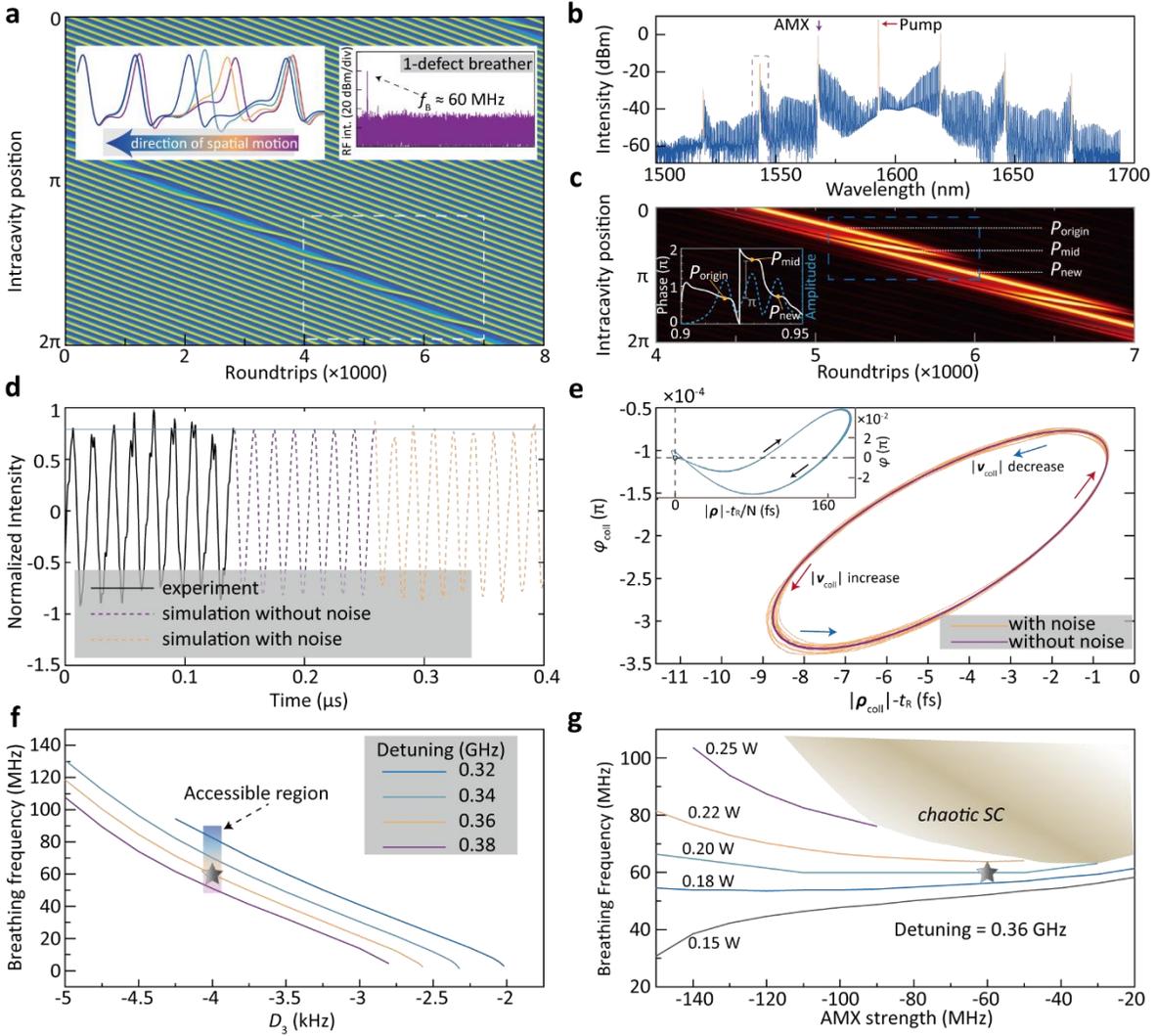

**Figure 3 | Physical mechanism of the breathing dynamics of soliton crystals (SCs). a**, Schematic of the 1-defect breather with a breathing period of ≈ 1,080 roundtrips, during which the soliton on one side of the defect moves to the other side. Left inset: The intracavity positions of the solitons near the defect are aligned and superimposed over multiple roundtrips to better illustrate the spatial motion. Right inset: The experimental RF spectrum with a breathing frequency close to the breathing frequency extracted from the cumulative intensity variation of the simulated crystal. **b**, The experimental OSA spectrum of a 1-defect breather with the strong comb lines indicating crystal spacing and structure marked in yellow. The pump and AMX positions are also shown. **c**, The simulated temporal waveform extracted from (**a**) after removing the corresponding prominent lines in the spectrum of 1-defect breather. Inset: the phase relation between three pulses. **d**, The experimental time-trace of breathing intensity measured by a high-speed PD and oscilloscope, shown in black. Corresponding simulations of the breathing time trace closely match the experiment and are plotted with and without noise in yellow and magenta respectively. **e**, The interaction plane limit cycle of the 1-defect breather plotted with and without noise in yellow and

magenta respectively corresponding to the plots in **(d)**. Inset: Simulated interaction plane limit cycle of all solitons in the crystal. **f,** The breathing frequency as a function of the third-order dispersion $D_3$ at different detunings. The shaded area is the experimentally accessible region, with the results closely matched by simulations. The grey star indicates the experimental conditions in **(d-e)**. **g**, The breathing frequency as a function of the AMX strength and pump power. The grey star indicates the conditions in **(d-e)**. Due to thermal dependence of the AMX strength, most area in **(g)** is experimentally accessible by varying the detuning or chip temperature.

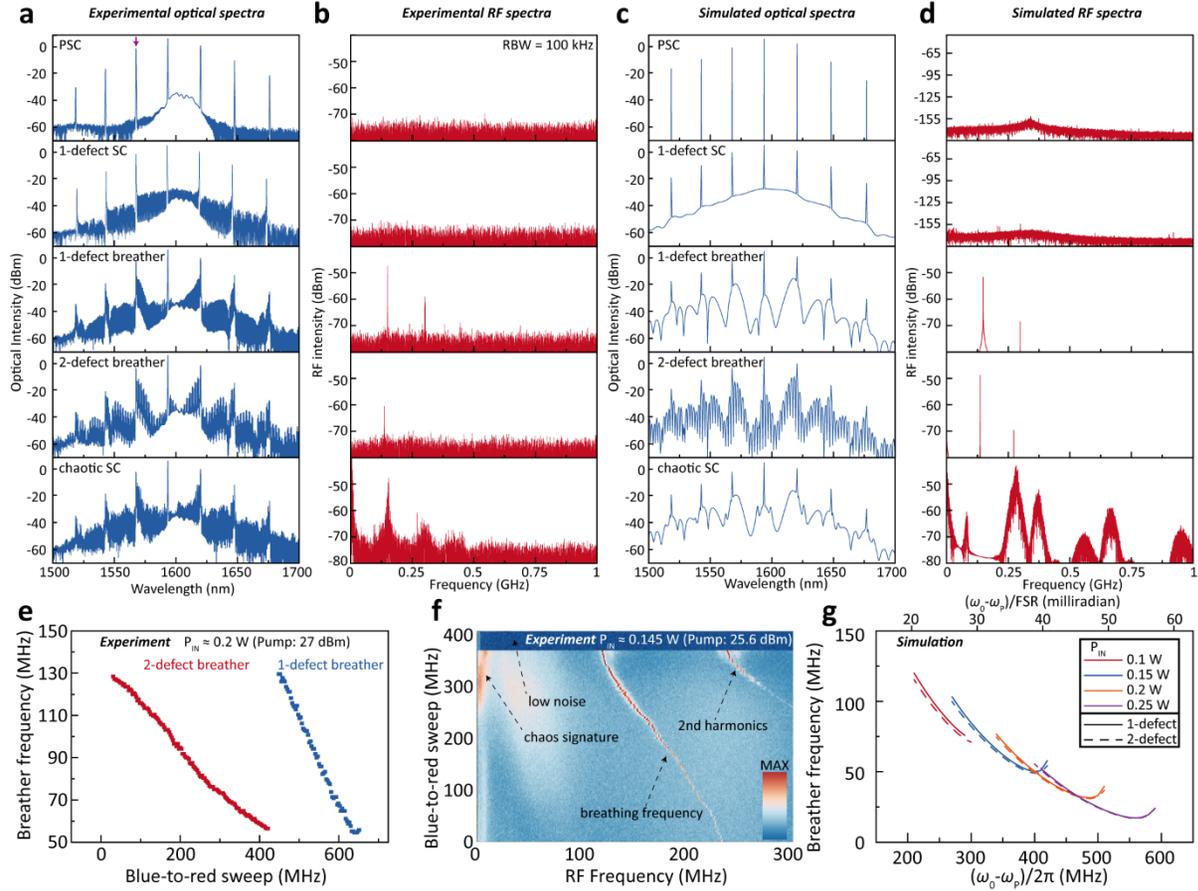

**Figure 4 | Measured-simulated optical and RF spectra, and breathing frequencies of the soliton crystals (SCs). a**, Experimental optical spectra of the PSC, 1-defect SC, 1-defect breather, 2-defect breather, and chaotic SC. The magenta arrow marks the dominant AMX point at $\mu = 48$. **b**, Experimental RF spectra corresponding to (**a**). The resolution bandwidth (RBW) is 100 kHz. **c**, Simulated optical spectra corresponding to (**a**). **d**, Simulated RF spectra corresponding to (**a**). **e**, Experimental evolution of the measured SC breathing frequency at 200 mW (pump = 27 dBm), tuned from the 2-defect breather (red) to the 1-defect breather (blue), with the transitions illustrated earlier in Fig. 2**c**. Red and blue blocks represent the recorded breathing frequencies. **f**, 2D map of detuning plotted against measured RF frequency. As we tune the pump to the red side, we clearly observe the transition from a single sharp breathing line, to a partially chaotic spectrum and then a sudden transition to a low noise state, in good agreement with Fig. 2**d** for a blue-to-red sweep from 1.80 to 2.15 GHz. **g**, Modeled breathing frequency of the 1- and 2-defect breathers as a function of detuning at different input powers.

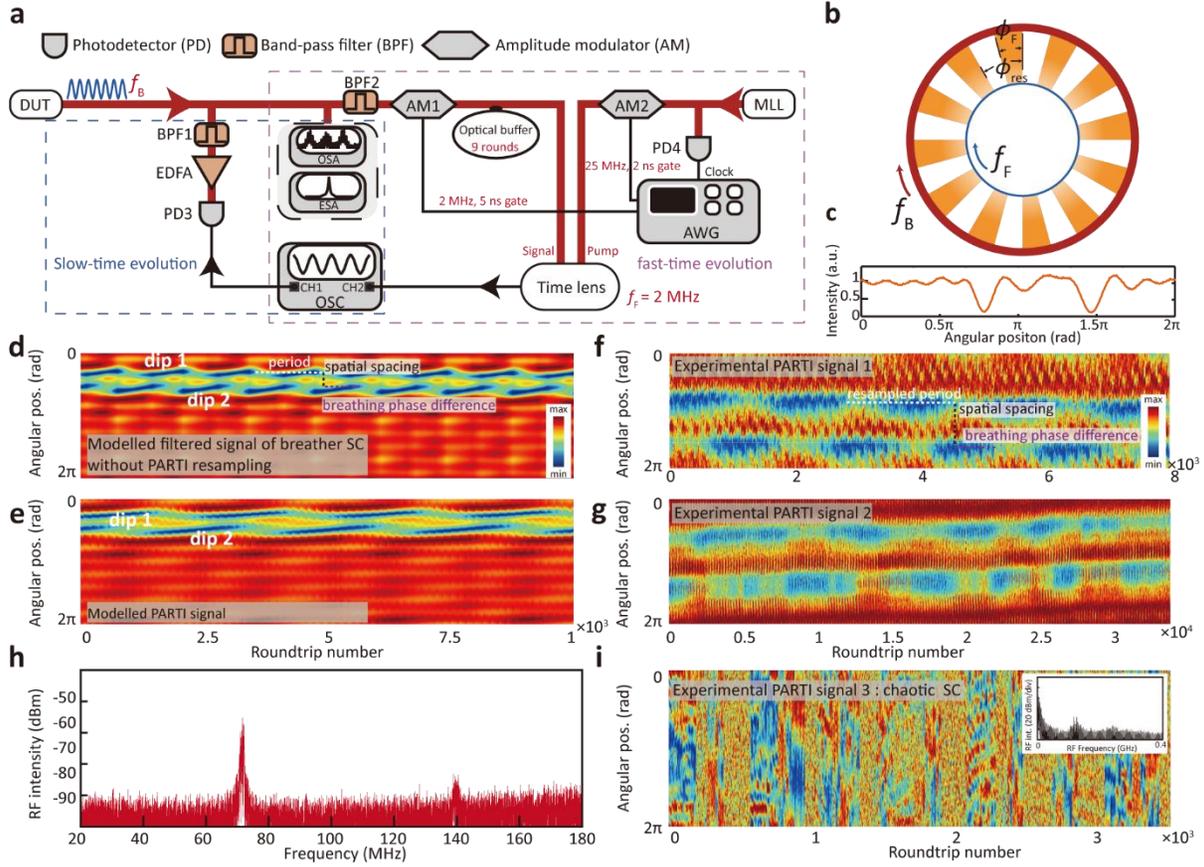

**Figure 5 | Temporal observation of the breathing dynamics of soliton crystals (SCs) with a 66 GHz Si₃N₄ microresonator. a**, Experimental setup for the simultaneous observation of the slow-time evolution (*blue dashed rectangular*) using oscilloscope directly and the fast-time evolution using an improved time-lens imaging system (*purple dashed rectangular*). Measurements with two different time scales are recorded separately. The SC microcombs are generated with a breathing angular frequency $f_B$. BPF1 and BPF2 have the different bandwidths. Optical buffer effectively extends the one-shot length of time-lens system to 2 ns. DUT, device under test; EDFA, erbium-doped fiber amplifier; ESA, electronic spectrum analyzer; OSA, optical spectrum analyzer; OSC, oscilloscope; MLL, mode-locked fiber laser; and AWG, arbitrary waveform generator. **b**, Schematic representation of the phase-differential sampling based on the stroboscopic effect. The blue circle represents the time lens system and the orange region $\phi_F$ represents the one-shot angular length. The red circle represents the period of the SC breather. We can use the stroboscopic effect to stitch the dynamics of breathers in a full period. $\phi_{res}$ is the accumulated phase difference between the SC breather and time-lens sampling after 500 ns. **c**, The simulated intensity distribution of 2-defect SC after BPF2 as a function of the fast time. One roundtrip is angularly normalized to be $2\pi$. The two defects are set to be temporally separated by 15 solitons. **d**, The simulated 2D evolution map of the intracavity intensity after a similar bandpass

filter as BPF2. Solitons near two defects move in different phases, creating two sequent dips in the spatio-temporal evolution map. Two dips are named "dip 1" and "dip 2" in the evolution map. The temporal period, spatial spacing and breathing phase difference of two dips are presented via white, black and purple dashed line, respectively. **e**, The 2D evolution map resampled from (**d**) to match the PARTI frame rate, where two dips are reconstructed via resampling and stitching different breathing periods. **f**, The 2D evolution map of the 2-defect breather recorded by our improved time-lens system, showing that the two solitons move in different phases as highlighted by the white dashed parallelogram. Intracavity one roundtrip is angularly normalized to be $2\pi$. The reconstructed temporal period, spatial spacing and breathing phase difference of two dips are presented via white, black and purple dashed line, respectively. **g**, The 2D evolution map of 2-defect breather with a decreased oscillation period. We clearly observe the drift in breather frequency due to noise in addition to the different breathing phases. **h**, The experimental radiofrequency (RF) spectra of the 2-defect breather. The occurrence of the second-harmonic peak near 140 MHz shows that the slow-time evolution is not a perfect sine wave. The broadening of the peak near 76 MHz explains the drift of breathing frequency in (**g**). **i**, Experimental panoramic-reconstruction temporal imaging (PARTI) trace of a chaotic SC with the inset showing the corresponding RF spectrum.

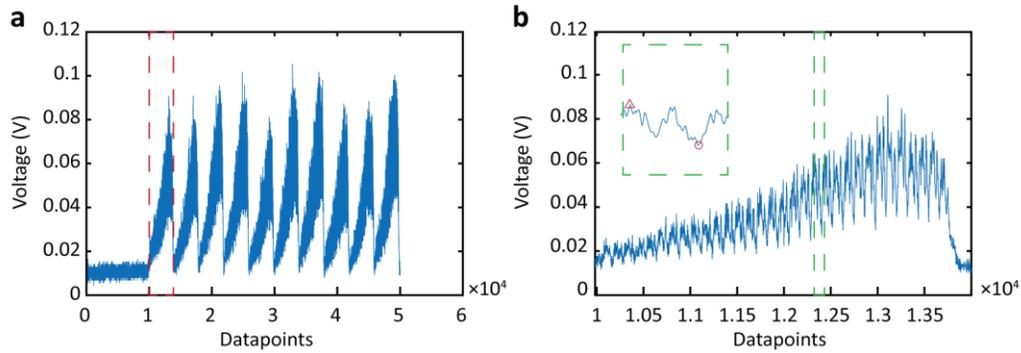

**Figure 6. a,** An example frame of PARTI results. **b,** The zoom-in of one replica outlined by the red dashed rectangle in (**a**). Inset: zoom-in of the signal in one round trip illustrated in the green dashed rectangle, where the triangle marks the voltage maximum and the circle marks the voltage minimum. The two major dips clearly have different voltages, implying that the two solitons move in different phases.